\documentclass[conference]{IEEEtran}
\IEEEoverridecommandlockouts
\usepackage{cite}
\usepackage{amsmath,amssymb,amsthm}
\usepackage{graphicx}
\usepackage{booktabs}
\usepackage{bm}
\usepackage{tikz}
\usetikzlibrary{arrows.meta,calc}
\newtheorem{theorem}{Theorem}
\newtheorem{proposition}{Proposition}

\usetikzlibrary{arrows.meta,calc,positioning}
\theoremstyle{remark}
\newtheorem{remark}{Remark}
\newcommand{\Mb}{\mathbf M}
\newcommand{\x}{\mathbf x}
\newcommand{\jw}{j\nu}
\newcommand{\kff}{k_{\mathrm{ff}}}
\newcommand{\hkff}{\hat k_{\mathrm{ff}}}
\newcommand{\Hinf}[1]{\|#1\|_\infty}

\begin{document}

\title{Course-Referenced Lateral Control of Trail-Following Convoys: String Stability and Adaptive Feedforward under Load Uncertainty}

\author{Nitesh Kumar$^{1}$, Sixu Li$^{2}$, and Swaroop Darbha$^{1}$%
\thanks{$^{1}$Nitesh Kumar and Swaroop Darbha are with the J.~Mike Walker '66 Department of Mechanical Engineering, Texas A\&M University, College Station, TX 77843, USA (e-mail: {\tt\small niteshk@tamu.edu; dswaroop@tamu.edu}).}%
\thanks{$^{2}$Sixu Li is with the Zachry Department of Civil and Environmental Engineering, Texas A\&M University, College Station, TX 77843, USA (e-mail: {\tt\small sixuli@tamu.edu}).}%
\thanks{Corresponding author: Sixu Li.}}

\maketitle

\begin{abstract}
Vehicles in a convoy can steer along the GNSS trails of the vehicles ahead when lane markings are unavailable, as on snow covered roads. Since each vehicle's tracking error enters the reference of the next, lateral errors can grow along the convoy. This paper presents a course referenced formulation of trail following in arc length, in which each follower takes the direction of its fitted trail as its heading reference. Error propagation then reduces to a single transfer function. We derive a condition under which combining the trails of the leader and the predecessor keeps the errors bounded for any convoy length, and show that following the predecessor alone lets the errors accumulate without bound, even without amplification, on roads that end turned or shifted sideways. Passenger and cargo loads and tire stiffness vary in practice, and we show that they enter the curvature feedforward through a single coefficient, whose mismatch leaves steady offsets on curves. We propose an adaptive law that estimates this coefficient on curves and removes the offsets. Simulations of an eleven vehicle convoy confirm the analysis.
\end{abstract}

\section{Introduction}

Lateral control of connected vehicle convoys matters where lane markings or road maps are unavailable or unreliable: in snow, on unmarked roads, and in emergency lane changes initiated by the convoy itself \cite{liu2024robust}. Each follower then receives GNSS waypoints from other vehicles, fits a path to them, and tracks the fitted path \cite{liu2021lateral,liu2024robust}. The tracking error of one vehicle becomes part of the reference of the next, so lateral errors can propagate along the convoy, as longitudinal spacing errors do in a platoon \cite{swaroop1996string,seiler2004disturbance}.

Several notions of \emph{lateral string stability} have been proposed. Hassanain et al.\ used $\mathcal H_\infty$ norms of the error transfer functions \cite{hassanain2020string}, and Alleleijn et al.\ introduced the idea earlier \cite{alleleijn2014lateral}. Papadimitriou et al.\ examined vision-based following and its Nyquist conditions \cite{papadimitriou2003autonomous}. Liu et al.\ proved that following the predecessor's trail alone amplifies low-frequency errors for every stabilizing gain, and proposed fitting a composite path to the trails of both the leader and the predecessor \cite{liu2024robust}. Somisetty and Darbha gave a sufficient condition for bounded errors with such a composite reference \cite{somisetty2024lateral}. Li et al.\ compared errors at the same arc length instead of the same time. They showed that onboard sensing with feedforward cannot be string stable, whereas learning from the predecessor over V2V can \cite{li2026lateral,li2026lateral1}.

This paper extends these results in two directions: to a course-referenced description of trail following that includes curvature feedforward, and to load uncertainty. We first fix the terms used throughout.
\begin{itemize}
\item The \emph{desired path} is the path the convoy should follow, such as the centreline of a lane. The leader, vehicle~$0$, tracks it using a map or its own sensors.
\item The \emph{trail} of a vehicle is the sequence of GNSS positions that it records and broadcasts as it drives, that is, samples of the path it actually traversed. A follower does not use its own trail.
\item Follower $i$ fits a smooth path to the trails of vehicles ahead: its predecessor's, the leader's, or a combination of the two with weight $\alpha\in[0,1]$ on the predecessor's trail and $1-\alpha$ on the leader's. It then steers to track this \emph{fitted path}.
\item The \emph{lateral error} $e_i$ of vehicle $i$ is the distance of its center of mass from the desired path, measured perpendicular to it. The \emph{offset} $r_i$ of the fitted path is defined in the same way, so $r_i$ is a weighted average of the lateral errors of the vehicles whose trails were fitted.
\end{itemize}
Because a trail is made of positions, the direction of a fitted path is the direction in which the vehicles ahead were \emph{moving}, their course angle. It is not the direction in which they were \emph{pointing}, their yaw angle. The two differ by the sideslip angle, which is nonzero whenever a vehicle corners (Fig.~\ref{fig:bicycle}). We call a formulation in which each follower takes the direction of its fitted path as its heading reference \emph{course-referenced}. The contributions of this paper are as follows.
\begin{enumerate}
\item \emph{Course-referenced model.} When vehicles are compared at the same position along the road, as in \cite{li2026lateral}, the direction of the fitted path relative to the desired path is the derivative of its offset $r_i$ with respect to distance. The fitted path is therefore described by one signal, $r_i$. The lateral error of each follower depends on it through a single transfer function $Q$, from the offset of the fitted path to the follower's lateral error (Section~\ref{sec:prop}), instead of through a $2\times2$ matrix acting on the lateral and heading errors of the vehicles ahead. The reference is also consistent in curves. A vehicle that corners exactly on the desired path lays a trail that lies on the path and points along it, so the follower sees no deviation, although that vehicle's yaw angle differs from the path direction by its sideslip angle.
\item \emph{An exact condition for bounded errors.} The lateral errors are bounded uniformly in convoy length if (and, for identical vehicles, only if) $\alpha\Hinf{Q}<1$, where $\Hinf{Q}$ is the peak gain of $Q$ over all spatial frequencies (Theorem~\ref{thm:lp}). The design of \cite{liu2024robust}, with its steering actuator and preview, and with the sideslip-compensated feedforward of Section~\ref{sec:ff}, satisfies $\alpha\Hinf{Q}\le0.67$ at 10--30~m/s, complementing the simulation results reported there.
\item \emph{Amplification versus accumulation.} A follower reproduces a constant offset of the path it follows, so $Q(0)=1$. Near zero frequency, $1-Q$ is proportional to the error in the curvature feedforward (Proposition~\ref{prop:lowfreq}). With feedforward of the fitted curvature, which extends the error-feedback law analyzed in \cite{liu2024robust}, low-frequency amplification under predecessor-only following can be avoided, and gains exist for which $\Hinf{Q}=1$. Even so, on any road that ends turned or shifted sideways, $Q(0)=1$ makes the errors injected by each vehicle accumulate without bound (Theorem~\ref{thm:pred}). Non-amplification is therefore not enough, and leader information is what bounds the errors.
\item \emph{Load uncertainty and adaptive feedforward.} Following the moment model of \cite{kumar2025robust}, loads and tire stiffness enter the sideslip-compensated feedforward only through one coefficient (Proposition~\ref{prop:ff}). We derive the resulting offsets and how they accumulate along the convoy, and give a gated, projected adaptive law for this coefficient. Its averaged dynamics are decoupled across vehicles (Proposition~\ref{prop:adapt}).
\end{enumerate}

\section{Model and controller}
\label{sec:model}

\subsection{Loaded bicycle model in path coordinates}

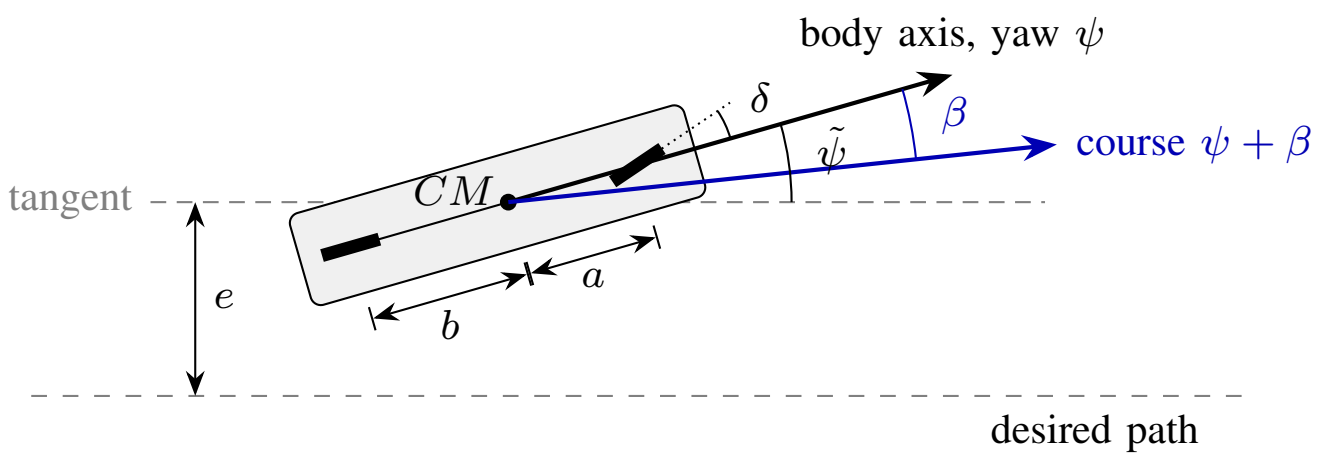
\begin{figure}[t]
\centering
\resizebox{\columnwidth}{!}{%
\begin{tikzpicture}[>=Stealth,font=\footnotesize]
\draw[gray,dashed] (-0.6,0) -- (7.6,0) node[pos=0.87,below,black]{desired path};
\coordinate (cm) at (2.6,1.3);
\draw[gray,densely dashed] (0.2,1.3) node[left,gray]{tangent} -- (6.2,1.3);
\draw[<->] (0.5,0) -- (0.5,1.3) node[midway,right]{$e$};
\begin{scope}[shift={(2.6,1.3)},rotate=16]
\draw[fill=gray!12,rounded corners=2pt] (-1.45,-0.32) rectangle (1.3,0.32);
\draw (-1.1,0)--(0.9,0);
\draw[line width=2.5pt] (-1.3,0)--(-0.9,0);
\begin{scope}[shift={(0.9,0)},rotate=18]\draw[line width=2.5pt] (-0.2,0)--(0.2,0);\draw[densely dotted] (0,0)--(0.75,0);\end{scope}
\draw[densely dotted] (0.9,0)--(1.7,0);
\draw (1.55,0) arc (0:18:0.65); \node at (1.82,0.21) {$\delta$};
\draw[|<->|] (-1.1,-0.5)--(0,-0.5) node[midway,below=-1pt]{$b$};
\draw[|<->|] (0,-0.5)--(0.9,-0.5) node[midway,below=-1pt]{$a$};
\end{scope}
\fill (cm) circle (1.6pt);
\node[above left= -4pt and -1pt] at (cm) {$CM$};
\draw[->,thick] (cm) -- ++(16:3.1) node[above]{body axis, yaw $\psi$};
\draw[->,thick,blue!70!black] (cm) -- ++(6:3.7) node[right]{course $\psi+\beta$};
\draw (cm) ++(0:1.9) arc (0:16:1.9); \node at ($(cm)+(9:2.2)$) {$\tilde\psi$};
\draw[blue!70!black] (cm) ++(6:2.75) arc (6:16:2.75); \node[blue!70!black] at ($(cm)+(11:3.05)$) {$\beta$};
\end{tikzpicture}}
\caption{Bicycle model. The heading error $\tilde\psi$ is the angle from the tangent of the desired path to the body axis. The sideslip angle $\beta$ is the angle from the body axis to the velocity of $CM$, here negative. The velocity points along the course, at $\tilde\psi+\beta$ from the tangent, which is the slope $e'$ of the vehicle's path.}
\label{fig:bicycle}
\end{figure}

Fig.~\ref{fig:bicycle} shows the bicycle model. Let $e$ be the lateral error of the unladen vehicle's center of mass ($CM$) from the desired path, and $\tilde\psi$ the heading error, the angle between the body axis and the tangent of the desired path. Write $\x=[e\ \ \tilde\psi]^T$. The desired path has curvature $\kappa$, the speed $V$ is constant, $a$ and $b$ are the distances from $CM$ to the axles, $\ell=a+b$, $\delta$ is the front-wheel steering angle, and $C_f,C_r$ are the axle cornering stiffnesses. The velocity of $CM$ has components $V$ along the body axis and $v_y$ across it. It makes the \emph{sideslip angle} $\beta=v_y/V$ with the body axis, so the vehicle moves along the course angle $\psi+\beta$ while pointing along its yaw angle $\psi$.

Following \cite{kumar2025robust}, passengers and cargo of masses $m_k$ at longitudinal positions $x_k\in[-d_r,d_f]$ relative to $CM$ enter only through the moments
\[
M_{11}=\textstyle\sum_km_k,\quad M_{12}=\sum_km_kx_k,\quad M_{22}=\sum_km_kx_k^2 .
\]
With $M_t=M_0+M_{11}$ and $I_t=I_0+M_{22}$, where $M_0$ and $I_0$ are the unladen mass and yaw inertia, the error dynamics are \cite{rajamani2012vehicle,kumar2025robust}
\begin{equation}
\Mb_L\ddot{\x}+\tfrac1V\mathbf C\dot{\x}+\mathbf L\x=\mathbf bC_f\delta-\mathbf f\kappa,
\label{eq:model}
\end{equation}
where $\Mb_L=\left[\begin{smallmatrix}M_t&M_{12}\\M_{12}&I_t\end{smallmatrix}\right]$, $\mathbf b=[1\ \ a]^T$, and
\begin{gather*}
\mathbf C=\begin{bmatrix}C_f+C_r&aC_f-bC_r\\aC_f-bC_r&a^2C_f+b^2C_r\end{bmatrix},\\
\mathbf L=\begin{bmatrix}0&-(C_f+C_r)\\0&-(aC_f-bC_r)\end{bmatrix},\quad
\mathbf f=\begin{bmatrix}M_tV^2+aC_f-bC_r\\M_{12}V^2+a^2C_f+b^2C_r\end{bmatrix}.
\end{gather*}
As is common \cite{rajamani2012vehicle}, we neglect the term in the curvature rate $\dot\kappa$, which comes from the angular acceleration of the path tangent; it vanishes on arcs and affects only the transients at curve entries. The off-diagonal moment $M_{12}$ represents the shift of the center of mass. It changes the understeer gradient, and a modest load behind the rear axle can make an understeering car neutral \cite{kumar2025robust}. The moments are realizable by nonnegative masses on $[-d_r,d_f]$ if and only if $\Mb_p=\left[\begin{smallmatrix}M_{11}&M_{12}\\M_{12}&M_{22}\end{smallmatrix}\right]\succeq0$ and $d_fd_rM_{11}-(d_r-d_f)M_{12}-M_{22}\ge0$ \cite{kumar2025robust}. The steering actuator is $\ddot\delta+2\zeta_\delta\omega_\delta\dot\delta+\omega_\delta^2\delta=\omega_\delta^2\delta_c$, with commanded angle $\delta_c$, $\zeta_\delta=0.41$ and $\omega_\delta=21.5$~rad/s identified in \cite{liu2024robust}.

\paragraph*{Arc length} Let $l$ be the arc length along the desired path and $(\cdot)'=d/dl$. At constant speed, $d/dt=V\,d/dl$, and \eqref{eq:model} becomes
\begin{equation}
V^2\Mb_L\x''+\mathbf C\x'+\mathbf L\x=\mathbf bC_f\delta-\mathbf f\kappa.
\label{eq:arc}
\end{equation}
The coefficients do not depend on $l$, so transfer functions in the Laplace variable $s$ of $l$, with spatial frequency $\nu$ at $s=\jw$, are well defined. A component at spatial frequency $\nu$ is a lateral wiggle of wavelength $2\pi/\nu$. Every vehicle is compared with its predecessor at the same place, so the time gap between them never enters the analysis \cite{li2026lateral}. A time-domain transfer function evaluated at $\omega=V\nu$ has the same $\mathcal H_\infty$ norm. In arc length, the lateral velocity is $\dot e=V(\tilde\psi+\beta)$ \cite{rajamani2012vehicle}, so
\begin{equation}
e'=\tilde\psi+\beta .
\label{eq:slope}
\end{equation}
The slope of a vehicle's path relative to the desired path is its course error, not its heading error.

\subsection{The fitted path and the measured errors}

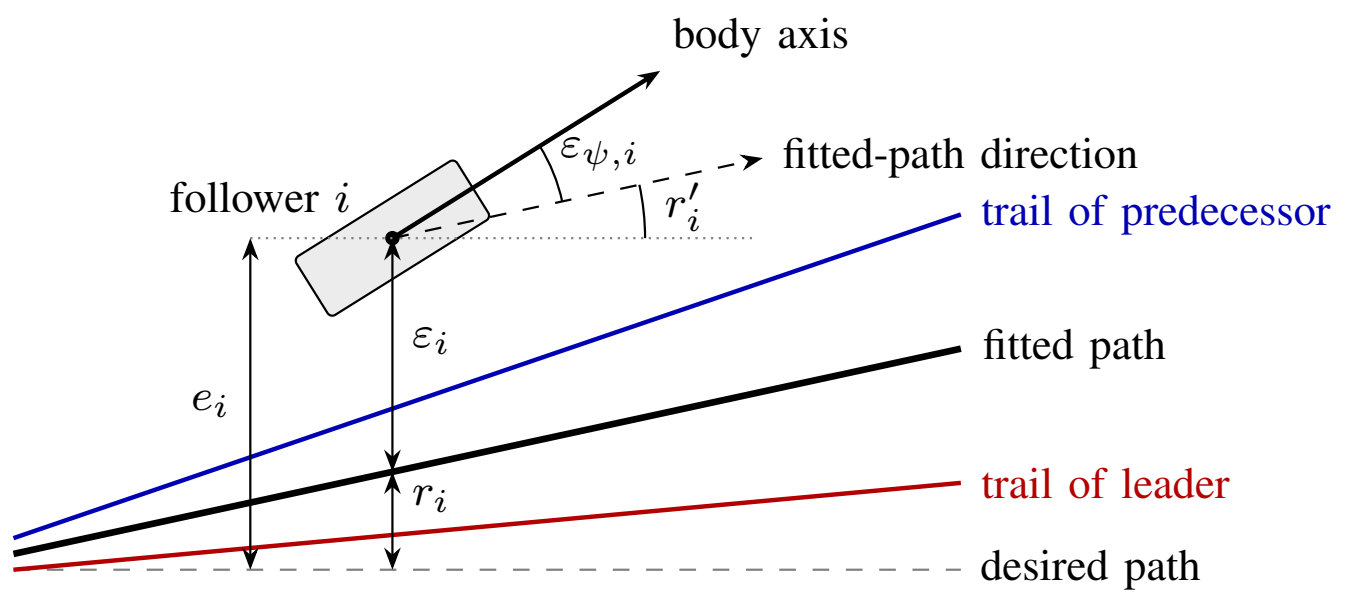
\begin{figure}[t]
\centering
\resizebox{\columnwidth}{!}{%
\begin{tikzpicture}[>={Stealth[length=4pt]},font=\footnotesize]
\draw[gray,dashed] (0,0) -- (6,0) node[right,black] {desired path};
\draw[red!70!black,thick] (0,0) -- (6,0.55) node[right] {trail of leader};
\draw[blue!70!black,thick] (0,0.2) -- (6,2.25) node[right] {trail of predecessor};
\draw[black,very thick] (0,0.1) -- (6,1.4) node[right] {fitted path};
\coordinate (cm) at (2.4,2.1);
\begin{scope}[shift={(cm)},rotate=32]
  \draw[fill=gray!15,rounded corners=1pt] (-0.6,-0.22) rectangle (0.6,0.22);
\end{scope}
\node[above left=1pt and 4pt] at (cm) {follower $i$};
\draw[gray,densely dotted] (cm) -- ++(0:2.3);
\draw[->,dashed] (cm) -- ++(12.2:2.4) node[right] {fitted-path direction};
\draw[->,thick] (cm) -- ++(32:2.0) node[above right=-2pt] {body axis};
\fill (cm) circle (1.3pt);
\draw (cm) ++(0:1.6) arc[start angle=0,end angle=12.2,radius=1.6];
\node at ($(cm)+(6.1:1.86)$) {$r_i'$};
\draw (cm) ++(12.2:1.1) arc[start angle=12.2,end angle=32,radius=1.1];
\node at ($(cm)+(22:1.42)$) {$\varepsilon_{\psi,i}$};
\draw[<->] (2.4,0) -- (2.4,0.62) node[pos=0.72,right] {$r_i$};
\draw[<->] (2.4,0.62) -- (cm);
\node[right] at (2.4,1.45) {$\varepsilon_i$};
\draw[gray,densely dotted] (1.5,2.1) -- (cm);
\draw[<->] (1.5,0) -- (1.5,2.1) node[midway,left] {$e_i$};
\end{tikzpicture}}
\caption{Errors measured by follower $i$ at its position $l$ along the road. The fitted path is offset by $r_i=\alpha e_{i-1}+(1-\alpha)e_0$ from the desired path (here $\alpha=0.5$), and its direction makes the angle $r_i'$ with the desired path. The follower measures its lateral error $\varepsilon_i=e_i-r_i$ and its heading error $\varepsilon_{\psi,i}$ relative to the fitted path.}
\label{fig:geometry}
\end{figure}

Each follower $i$ fits a circular-arc spline to the waypoints of the leader and of its predecessor, combined with weights $1-\alpha$ and $\alpha$ \cite{liu2024robust}.\footnote{In \cite{liu2024robust} the weight $\alpha$ multiplies the leader's fit; here it multiplies the predecessor's, as in \cite{somisetty2024lateral}. Both papers use $\alpha=0.5$.} With $\alpha=1$, a follower uses its predecessor's trail alone. Least squares is linear in the data. To first order, the fitted path is therefore offset from the desired path by
\begin{equation}
r_i(l)=\alpha e_{i-1}(l)+(1-\alpha)e_0(l),\qquad r_1=e_0,
\label{eq:ref}
\end{equation}
as shown in Fig.~\ref{fig:geometry}. The follower measures three errors relative to the fitted path: its lateral error, its heading error, and the rate of change of the heading error,
\begin{equation}
\varepsilon_i=e_i-r_i,\quad
\varepsilon_{\psi,i}=\tilde\psi_i-r_i',\quad
\varepsilon_{\omega,i}=V\bigl(\tilde\psi_i'-r_i''\bigr).
\label{eq:errors}
\end{equation}
The heading reference is $r_i'$, the direction of the fitted path, which is what the fit provides. By \eqref{eq:slope}, it is the course error of the vehicles ahead, not their heading error. In steady cornering on the desired path, a vehicle has $e'=0$ while $\tilde\psi=-\beta\neq0$. Such a vehicle lays a trail that is tangent to the desired path, so the follower sees no heading deviation. A reference built from that vehicle's yaw angle would instead show a deviation equal to its sideslip angle. The third error, $\varepsilon_{\omega,i}$, is the yaw-rate error relative to the rate at which the fitted path turns.

\subsection{Controller and sideslip-compensated feedforward}
\label{sec:ff}

The steering command is \cite{liu2024robust}
\begin{equation}
\delta_{c,i}=\hkff\,\hat\kappa_i-k_e\varepsilon_i-k_\psi\varepsilon_{\psi,i}-k_\omega\varepsilon_{\omega,i}.
\label{eq:law}
\end{equation}
The last three terms are feedback with gains $k_e,k_\psi,k_\omega$. The first is feedforward: $\hat\kappa_i$ is the curvature of the fitted path, and $\hkff$ is a gain chosen below. The fit averages the curvature of the trail, $\kappa+r_i''$, over a preview window of length $L_p=0.8V$ \cite{liu2024robust}, so that
\[
\hat\kappa_i=W(s)\,(\kappa+r_i''),\qquad W(s)=\frac{e^{sL_p}-1}{sL_p}.
\]
The vehicles ahead have already passed through the window, so this preview is realizable.

\paragraph*{Why the feedforward gain has the form $\theta_1+\theta_2V^2$} On an arc of curvature $\kappa$, the vehicle needs a constant steering angle $\delta^\ast$ to stay on the arc. This angle has two parts. The first is the geometric (Ackermann) angle $\ell\kappa$. The second comes from the tire slip angles needed to produce the centripetal force, and is proportional to the lateral acceleration $V^2\kappa$. Moreover, in steady cornering the body axis is not tangent to the path: the heading error settles at $\tilde\psi_{ss}=-\beta_{ss}\neq0$. The heading feedback $-k_\psi\tilde\psi$ would steer against this harmless heading error, and the position feedback would then need a nonzero lateral error to cancel it. The feedforward must therefore supply $\delta^\ast+k_\psi\tilde\psi_{ss}$. Both terms are linear in $\kappa$, with a part independent of speed and a part proportional to $V^2$. The required gain is therefore $\kff=\theta_1+\theta_2V^2$, where $\theta_1$ collects the geometric terms and $\theta_2$ the terms proportional to lateral acceleration, which carry the mass and tire parameters. The next result gives both.

\begin{proposition}[Feedforward for a loaded vehicle]
\label{prop:ff}
On an arc of constant curvature $\kappa\neq0$, the closed loop \eqref{eq:arc}, \eqref{eq:law} with $r_i\equiv0$ and stabilizing gains has zero steady-state lateral error if and only if $\hkff=\kff:=\theta_1+\theta_2V^2$, where
\begin{gather}
\theta_1=\ell-k_\psi b,\nonumber\\
\theta_2=\frac{M_t}{\ell}\Bigl(\frac b{C_f}-\frac{(1-k_\psi)a}{C_r}\Bigr)+\frac{M_{12}}{\ell}\Bigl(\frac1{C_f}+\frac{1-k_\psi}{C_r}\Bigr).
\label{eq:theta}
\end{gather}
The steady heading error is independent of $\hkff$: $\tilde\psi_{ss}=-\kappa\bigl[b-(aM_t-M_{12})V^2/(C_r\ell)\bigr]$. For any $\hkff$, the steady lateral error is $e_{ss}=(\hkff-\kff)\kappa/k_e$.
\end{proposition}

\begin{proof}
In steady state on the arc, $\x'=\x''=0$, $\delta=\delta_c$ and $\hat\kappa_i=\kappa$, so \eqref{eq:arc} reduces to $\mathbf L\x=\mathbf bC_f\delta-\mathbf f\kappa$ and \eqref{eq:law} to $\delta=\hkff\kappa-k_ee-k_\psi\tilde\psi$. The first column of $\mathbf L$ is zero, so $e$ enters only through $\delta$. Subtracting $a$ times the first row from the second eliminates $\delta$ and leaves $\ell C_r\tilde\psi=-\bigl[b\ell C_r+(M_{12}-aM_t)V^2\bigr]\kappa$, which gives $\tilde\psi_{ss}$ for any $\hkff$. The first row then gives the unique steering angle $\delta^\ast$ that holds the arc. Equating it with the control law,
\[
k_e e_{ss}=\hkff\kappa-\bigl(\delta^\ast+k_\psi\tilde\psi_{ss}\bigr)=(\hkff-\kff)\kappa,
\]
where $\kff\kappa:=\delta^\ast+k_\psi\tilde\psi_{ss}$. Hence $e_{ss}=0$ if and only if $\hkff=\kff$. Substituting $\delta^\ast$ and $\tilde\psi_{ss}$ and collecting powers of $V$ gives \eqref{eq:theta}.
\end{proof}

The coefficient $\theta_1$ depends only on geometry and the gain $k_\psi$, and is known. The coefficient $\theta_2$ depends on the load and the tires. We therefore use
\[
\hkff=\theta_1+\hat\theta_2V^2,
\]
where $\hat\theta_2$ is either the nominal (unladen) value from \eqref{eq:theta} or the adaptive estimate of Section~\ref{sec:adapt}. Compared with the feedforward $\ell\kappa+K_{us}V^2\kappa/g$ of \cite{liu2024robust}, \eqref{eq:theta} adds the terms in $k_\psi$, which compensate the heading feedback acting on the steady sideslip, and accounts for the load. For $M_{12}=0$, it agrees with the feedforward gain of \cite{li2026lateral}. Three observations follow:
\begin{itemize}
\item $\theta_1$ depends only on geometry and gains;
\item the yaw inertia $M_{22}$ does not enter at all;
\item all uncertainty in load and tires is collected in the single coefficient $\theta_2$, which is affine in $(M_{11},M_{12})$.
\end{itemize}

\section{Lateral string stability in arc length}
\label{sec:string}

\subsection{Error propagation and the transfer function $Q$}
\label{sec:prop}

Take Laplace transforms in $l$ with zero initial conditions. Since $d/dt=V\,d/dl$, the actuator is
\[
A(s)=\frac{\omega_\delta^2}{V^2s^2+2\zeta_\delta\omega_\delta Vs+\omega_\delta^2},\qquad \hat\delta=A(s)\,\hat\delta_c,
\]
and \eqref{eq:arc} reads $\mathbf P(s)\x=\mathbf bC_f\delta-\mathbf f\kappa$ with $\mathbf P(s)=V^2\Mb_Ls^2+\mathbf Cs+\mathbf L$. Substituting the errors \eqref{eq:errors} and $\hat\kappa_i$ into \eqref{eq:law} separates the command into feedback on the follower's own errors and the terms contributed by the fitted path:
\begin{gather}
\delta_{c,i}=-\mathbf k(s)\x_i+q(s)\,r_i+\hkff W(s)\,\kappa,\nonumber\\
\mathbf k(s)=\bigl[k_e\ \ k_\psi+k_\omega Vs\bigr],\nonumber\\
q(s)=k_e+k_\psi s+\bigl(k_\omega V+\hkff W(s)\bigr)s^2.
\label{eq:qsplit}
\end{gather}
The fitted path enters only through its offset $r_i$. Its slope and curvature, $r_i'$ and $r_i''$, are the factors $s$ and $s^2$ in $q(s)$. The closed loop is
\begin{multline*}
\bigl[\mathbf P(s)+C_fA(s)\,\mathbf b\,\mathbf k(s)\bigr]\x_i\\
=C_fA(s)\,\mathbf b\,\bigl[q(s)r_i+\hkff W(s)\kappa\bigr]-\mathbf f\kappa.
\end{multline*}
The feedback matrix $C_fA\mathbf b\mathbf k$ has rank one. By the matrix determinant lemma, the closed-loop determinant is
\begin{multline}
\Delta_{cl}(s)=s^2D(s)\\
+C_fA(s)\bigl[k_eN_e(s)+(k_\psi+k_\omega Vs)N_\psi(s)\bigr],
\label{eq:dcl}
\end{multline}
and $[\mathbf P+C_fA\mathbf b\mathbf k]^{-1}\mathbf b=\operatorname{adj}(\mathbf P)\mathbf b/\Delta_{cl}$. Here $\det\mathbf P(s)=s^2D(s)$, $\operatorname{adj}(\mathbf P)\mathbf b=[N_e\ \ N_\psi]^T$, and
\begin{gather*}
N_e(s)=V^2(I_t-aM_{12})s^2+b\ell C_rs+\ell C_r,\\
N_\psi(s)=s\bigl[V^2(aM_t-M_{12})s+\ell C_r\bigr],\\
D(s)=d_2s^2+d_1s+d_0,\quad d_2=V^4(M_tI_t-M_{12}^2),
\end{gather*}
with $d_1=V^2\bigl[M_t(a^2C_f+b^2C_r)+I_t(C_f+C_r)-2M_{12}(aC_f-bC_r)\bigr]$ and $d_0=C_fC_r\ell^2-V^2\bigl[M_t(aC_f-bC_r)-M_{12}(C_f+C_r)\bigr]$.
The gains are stabilizing when the polynomial $(V^2s^2+2\zeta_\delta\omega_\delta Vs+\omega_\delta^2)\Delta_{cl}(s)$ has all its roots in the open left half-plane. The first component of $\x_i$ gives
\begin{gather}
e_i=Q(s)\,r_i+w_i,\label{eq:prop}\\
Q(s)=T(s)\,q(s),\qquad T(s)=\frac{C_fA(s)N_e(s)}{\Delta_{cl}(s)}.\label{eq:Q}
\end{gather}
Here $T$ is the closed-loop transfer function from a steering command added at the actuator input to the lateral error. The term $w_i$ is the response to the road curvature: the error vehicle $i$ would make if it tracked the desired path itself. For the leader, $r_0=0$ and $e_0=w_0$. Since $N_\psi(0)=0$ and $A(0)=1$, $T(0)=1/k_e$.

\paragraph*{Meaning of $Q$} $Q$ maps the offset of the fitted path to the follower's lateral error, both measured from the desired path at the same place. At spatial frequency $\nu$, a wiggle of the fitted path of wavelength $2\pi/\nu$ and amplitude $1$ reappears in the follower's path with amplitude $|Q(\jw)|$. If $|Q|<1$ the wiggle is attenuated; if $|Q|>1$ it is amplified. Combining \eqref{eq:prop} with \eqref{eq:ref} gives the propagation along the convoy,
\[
e_i=Q\bigl[\alpha e_{i-1}+(1-\alpha)e_0\bigr]+w_i .
\]
This relation is scalar: one transfer function acting on lateral errors. In the vector formulation of \cite{liu2024robust,somisetty2024lateral}, the corresponding map is a rank-one $2\times2$ matrix acting on lateral and heading errors. The course reference reduces it to $Q$, because the heading reference $r_i'$ is the derivative of $r_i$ and is already contained in $q(s)$.

\begin{proposition}[Low-frequency structure]
\label{prop:lowfreq}
For every stabilizing gain and every $\hkff$,
\begin{equation}
Q(0)=1,\quad Q'(0)=0,\quad 1-Q(s)=-\frac{\hkff-\kff}{k_e}\,s^2+O(s^3).
\label{eq:lowfreq}
\end{equation}
Consequently $|Q(\jw)|=1-(\hkff-\kff)\nu^2/k_e+O(\nu^3)$. If $\hkff<\kff$, then $\Hinf{Q}>1$.
\end{proposition}

\begin{proof}
Use the final-value theorem with $\kappa=0$ and references $r=c$, $r=cl$ and $r=cl^2/2$, whose transforms are $c/s$, $c/s^2$ and $c/s^3$. In each case $\varepsilon=e-r=(Q-1)r$.
\begin{itemize}
\item A constant offset of the trail is followed exactly, since $\mathbf L$ annihilates $[1\ 0]^T$. Hence $Q(0)=1$.
\item A straight trail at a small angle $c$ is followed with $e'=\tilde\psi=c$ and $\delta=0$, because a vehicle driving straight has no slip. All three errors in \eqref{eq:errors} vanish, so $Q'(0)=0$.
\item A trail of constant curvature $c$ requires, by Proposition~\ref{prop:ff}, the steering $\kff c$, while \eqref{eq:law} supplies $\hkff W(0)c=\hkff c$. Hence $\varepsilon_{ss}=(\hkff-\kff)c/k_e=-c\lim_{s\to0}(1-Q)/s^2$.
\end{itemize}
\end{proof}

\begin{remark}
Without curvature feedforward ($\hkff=0$), $\Hinf{Q}>1$ whenever $\kff>0$, which holds for all gains considered here. This agrees with the double-integrator obstruction established for error feedback in \cite{liu2024robust,papadimitriou2003autonomous}. Proposition~\ref{prop:lowfreq} adds that, once the fitted curvature is fed forward, the obstruction is governed by the feedforward error: an underestimated feedforward makes the follower lag inside the curves of the trail. Exact feedforward makes $1-Q=O(s^3)$, and the obstruction disappears.
\end{remark}

\subsection{Leader and predecessor information}

\begin{theorem}[Bounded errors]
\label{thm:lp}
Let $0\le\alpha<1$, let $\bar q=\sup_i\Hinf{Q_i}$, and let $\bar w=\sup_i\|w_i\|_2$.
\begin{enumerate}
\item[(a)] If $\rho:=\alpha\bar q<1$, then for every convoy length
\[
\sup_i\|e_i\|_2\le\bar q\|e_0\|_2+\bar w+\frac{(1-\alpha)\bar q\,\|e_0\|_2+\bar w}{1-\rho}.
\]
\item[(b)] Let the vehicles be identical and $w_i=0$ for $i\ge1$. If $\alpha|Q(\jw_0)|>1$ for some $\nu_0\neq0$, there is a leader error $e_0\in\mathcal L_2$ for which $\sup_i\|e_i\|_2=\infty$.
\end{enumerate}
\end{theorem}

\begin{proof}
(a) For $i\ge2$, \eqref{eq:prop} and \eqref{eq:ref} give $\|e_i\|\le\rho\|e_{i-1}\|+(1-\alpha)\bar q\|e_0\|+\bar w$, and $\|e_1\|\le\bar q\|e_0\|+\bar w$. Iterate.
(b) For $i\ge2$, $e_i-e_{i-1}=\alpha Q(e_{i-1}-e_{i-2})$ with $e_1-e_0=(Q-1)e_0$, so $e_i-e_{i-1}=(\alpha Q)^{i-1}(Q-1)e_0$. Since $\alpha|Q(\jw_0)|>1\ge\alpha$, $Q(\jw_0)\neq1$. By continuity, choose $e_0$ with spectrum in a neighbourhood where $\alpha|Q|>1+\eta$ and $Q\neq1$. The differences grow geometrically.
\end{proof}

The condition $\alpha\bar q<1$ is thus necessary and sufficient for identical vehicles. Table~\ref{tab:gains} evaluates it for the gains $K_A=(k_e,k_\psi,k_\omega)=(0.06,0.96,0.08)$ of \cite{liu2024robust}, with the actuator,  the preview filter, and the nominal sideslip-compensated feedforward, over loads of up to $300$~kg at nine positions between the bumpers. With $\alpha=0.5$, $\alpha\bar q\le0.67$. The largest admissible weight is $\alpha^\star=1/\bar q=0.74$ at $10$~m/s and $0.85$ at $30$~m/s. The largest values occur near the lightly damped actuator mode. Theorem~\ref{thm:lp} thus provides an analytic certificate that complements the simulations of \cite{liu2024robust} for $\alpha=0.5$.

\begin{table}[h]
\centering
\caption{String gain $\Hinf{Q}$ with the steering actuator. Loads: up to $300$~kg at nine positions; ``nominal'' means unladen $\hat\theta_2$ in the sideslip-compensated feedforward~\eqref{eq:theta}.}
\label{tab:gains}
\setlength{\tabcolsep}{4pt}
\begin{tabular}{@{}llccc@{}}
\toprule
Gains & Model / feedforward & 10 m/s & 20 m/s & 30 m/s\\
\midrule
$K_A$ & course, no FF & 1.45 & 1.37 & 1.26\\
$K_A$ & course, nominal FF, worst load & 1.34 & 1.27 & 1.17\\
$K_B$ & course, no FF & 1.15 & 1.25 & 1.54\\
$K_B$ & course, nominal FF, worst load & 1.001 & 1.007 & 1.029\\
$K_B$ & course, exact FF, every load & 1.000 & 1.000 & 1.000\\
\bottomrule
\end{tabular}
\end{table}

\begin{figure}[h]
\centering
\includegraphics[width=\columnwidth]{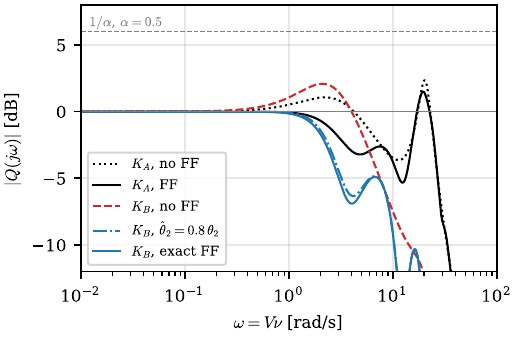}
\caption{$|Q|$ at $25$~m/s, unladen, with the steering actuator. With exact feedforward, $K_B$ has $|Q|\le1$ at every frequency. Underestimating $\theta_2$ by $20\%$ restores a low-frequency peak above $0$~dB, as Proposition~\ref{prop:lowfreq} predicts. The peaks near $20$~rad/s come from the actuator mode.}
\label{fig:q}
\end{figure}

\subsection{Predecessor information only}

With $\alpha=1$, Proposition~\ref{prop:lowfreq} leaves room for non-amplification. A search over gains, with the actuator, the preview filter and the load set of Table~\ref{tab:gains}, found $K_B=(0.03,0.5,0.02)$. With exact feedforward, $\Hinf{Q}=1$ at $10,15,\dots,30$~m/s for every load in the set (Table~\ref{tab:gains}, Fig.~\ref{fig:q}). The closed-loop damping ratio is at least $0.35$. Feedforward of the fitted curvature therefore makes non-amplifying predecessor-only following possible. The next result shows that non-amplification alone does not bound the errors.

\begin{theorem}[Accumulation]
\label{thm:pred}
Let $\alpha=1$, let the vehicles be identical, and let $w_i=w$ for all $i\ge0$, with transform $\mathcal W(s)$. Let $k\ge0$ and $m\ge2$ be integers with $k\le m-1$ such that $1-Q(s)=O(s^m)$ and $\mathcal W(s)=w_ks^k+O(s^{k+1})$ with $w_k\neq0$. Then, for every stabilizing gain, there is $c>0$ such that $\|e_N\|_2\ge cN^{1-(2k+1)/(2m)}$ for all sufficiently large $N$.
\end{theorem}

\begin{proof}
From \eqref{eq:prop}, $e_N=\sum_{j=0}^NQ^jw$, whose transform is $E_N=\sum_{j=0}^NQ^j\mathcal W$. There are $C,\nu_1>0$ such that $|Q(\jw)-1|\le C\nu^m$ and $|\mathcal W(\jw)|\ge|w_k|\nu^k/2$ for $0\le\nu\le\nu_1$. Let $\nu_N:=(4CN)^{-1/m}$, and let $N$ be large enough that $\nu_N\le\nu_1$. For $\nu\le\nu_N$ and $j\le N$, $|Q^j-1|\le jC\nu^m(1+C\nu^m)^{j}\le e^{1/4}/4<1/2$, so $|E_N(\jw)|\ge(N+1)|w_k|\nu^k/4$. Parseval's theorem gives $\|e_N\|_2^2\ge(N+1)^2|w_k|^2\nu_N^{2k+1}/(16\pi(2k+1))$, which grows as $N^{2-(2k+1)/m}$.
\end{proof}

By Proposition~\ref{prop:lowfreq}, $m=2$ when $\hkff\neq\kff$, and $m\ge3$ when $\hkff=\kff$. The transform $\mathcal W$ is the transfer function from road curvature to error times the transform of $\kappa$. At $s=0$, the first factor is the steady offset per unit curvature, $(\hkff-\kff)/k_e$ by Proposition~\ref{prop:ff}, and the second is the net heading change $\int\kappa\,dl$ of the road. With a feedforward error, $k=0$ on a road with a net heading change, and $k=1$ on a road with no net heading change but a net lateral shift, such as that of Section~\ref{sec:sim}. With exact feedforward, generically $k=1$ and $k=2$ in these two cases. In all four cases $k\le m-1$, and the errors grow without bound.

Theorems~\ref{thm:lp} and~\ref{thm:pred} support the conclusion of \cite{liu2024robust,somisetty2024lateral} that leader information is needed, and add a mechanism that persists with feedforward. Because $Q(0)=1$, each vehicle passes on what it inherits and adds its own error, even when there is no amplification. A weight $\alpha<1$ on the predecessor makes the low-frequency gain $\alpha<1$ and bounds the sum. The learn-from-predecessor scheme of \cite{li2026lateral} avoids this accumulation in a different way: each vehicle corrects toward the common desired path.

\section{Load uncertainty and adaptive feedforward}
\label{sec:adapt}

\subsection{Offsets and their accumulation}

Let $\tilde\theta_{2,i}=\hat\theta_{2,i}-\theta_{2,i}$. On an arc of curvature $\kappa$, Proposition~\ref{prop:ff} gives vehicle $i$ the steady error $d_i=\tilde\theta_{2,i}V^2\kappa/k_e$ relative to its own trail. Since $Q_i(0)=1$, \eqref{eq:prop} at steady state reads $e_i=\alpha e_{i-1}+(1-\alpha)e_0+d_i$. The offsets therefore add up:
\begin{itemize}
\item with $\alpha=1$, $e_N=\sum_{i\le N}d_i$, which grows linearly when the loads have the same sign of mismatch;
\item with $\alpha<1$, $|e_i-e_0|\le\max_j|d_j|/(1-\alpha)$.
\end{itemize}
Since $\theta_2$ is affine in $(M_{11},M_{12})$ and $-d_rM_{11}\le M_{12}\le d_fM_{11}$, its range over loads of at most $M_{\max}$ is attained at no load or at the whole load at a bumper. Interval bounds on $C_f$ and $C_r$ then give an interval $\Theta=[\theta_{\min},\theta_{\max}]$. For $M_{\max}=400$~kg and tire stiffness between $80\%$ and $110\%$ of nominal, this gives $\Theta=[-0.2,\,4.1]\times10^{-3}$~s$^2$/m for $K_B$, against the nominal $1.5\times10^{-3}$. At $25$~m/s on a $250$~m radius, a $20\%$ error in $\theta_2$ gives $d_i\approx2.5$~cm with $K_B$.

\subsection{Adaptive law}

The offset $d_i$ is measured directly as $\varepsilon_i$ once the vehicle has settled on an arc, and its sign follows that of $\tilde\theta_{2,i}$. This suggests the normalized gradient law
\begin{equation}
\dot{\hat\theta}_{2,i}=\mathrm{Proj}_\Theta\Bigl\{-\bar\gamma\,g_i(t)\,\frac{k_e\phi_i\,\varepsilon_i}{\phi_i^2+\phi_0^2}\Bigr\},\qquad\phi_i=V^2\hat\kappa_i,
\label{eq:adapt}
\end{equation}
where $\mathrm{Proj}_\Theta$ is the standard projection onto $\Theta$ \cite{ioannou1996robust}, and $\phi_0>0$ avoids division by zero. The gate $g_i\in\{0,1\}$ is on when the fitted path is an arc, $|\hat\kappa_i|>\kappa_{\min}$, whose curvature has stayed within $5\%$ for the last $T_g$ seconds and over the preview window. The arc-spline fit of \cite{liu2024robust} already classifies segments as arcs or lines. The gate excludes the transients at curve entries, which would otherwise bias the estimate. Projection keeps $\hat\theta_{2,i}\in\Theta$ at all times, and therefore $\hkff$ bounded \cite{ioannou1996robust}.

\begin{proposition}[Convergence]
\label{prop:adapt}
Let vehicle $i$ drive on an arc with $g_i=1$, $\phi_i=\phi\neq0$ constant, and a reference $r_i$ that has settled to a steady arc. Then there is $\gamma^\star>0$ such that, for $0<\bar\gamma<\gamma^\star$, the equilibrium $\hat\theta_{2,i}=\theta_{2,i}$, $\varepsilon_i=0$ is exponentially stable. The estimation error decays as $\exp(-\bar\gamma\phi^2t/(\phi^2+\phi_0^2))$ to first order in $\bar\gamma$, and the rate does not depend on the position $i$ in the convoy.
\end{proposition}

\begin{proof}
With $\bar\gamma$ as the small parameter, \eqref{eq:arc}, \eqref{eq:law} and \eqref{eq:adapt} form a singularly perturbed system, in which $\hat\theta_{2,i}$ is the slow state. The boundary-layer system is the closed loop with $\hat\theta_{2,i}$ frozen, which is exponentially stable by the choice of gains, and its equilibrium is the steady arc. By Proposition~\ref{prop:ff} and $Q_i(0)=1$, the steady $\varepsilon_i$ equals $\tilde\theta_{2,i}\phi/k_e$, independent of $r_i$. The reduced system is therefore
\[
\dot{\tilde\theta}_{2,i}=-\bar\gamma\frac{\phi^2}{\phi^2+\phi_0^2}\tilde\theta_{2,i},
\]
which is exponentially stable. Both systems are linear in $\tilde\theta_{2,i}$ at equilibrium, and the result follows from the exponential stability theorem for singularly perturbed systems \cite[Thm.~11.4]{khalil2002nonlinear}.
\end{proof}

The reduced dynamics of different vehicles are decoupled, because a follower's steady error depends only on its own feedforward error. On straight roads $\phi_i=0$, so nothing is learned, but nothing needs to be. The estimate carries over to arcs of other radii and to other speeds, because $\theta_2$ multiplies $V^2\kappa$. Integral action on $\varepsilon_i$ would also remove the offset, but it would add a pole at the origin to the loop and change $Q$. The adaptive feedforward leaves $Q$ unchanged once it has converged, and keeps what it has learned from one curve to the next.

\section{Simulations}
\label{sec:sim}

An eleven-vehicle convoy (leader and ten followers) drives at $25$~m/s along the following desired path:
\begin{itemize}
\item $200$~m straight;
\item a $3.6$~m lane change made of four $30$~m arcs of radius $500$~m;
\item $180$~m straight;
\item a $400$~m left arc and, after $100$~m of straight road, a $400$~m right arc, both of radius $250$~m;
\item a final straight.
\end{itemize}
Each curvature step is replaced by a $10$~m clothoid. All numerical results, including those of Sections~\ref{sec:string} and~\ref{sec:adapt}, use the vehicle of \cite{liu2024robust}: $M_0=1896$~kg, $I_0=3803$~kg\,m$^2$, $C_f=4.0\times10^5$, $C_r=3.82\times10^5$~N/rad, $a=1.27$~m, $b=1.58$~m, with overhangs of $0.9$~m and $1.0$~m. Each vehicle is the loaded model \eqref{eq:model} with the actuator, and the system is integrated in arc length as described in Section~\ref{sec:model}. All vehicles carry two rear passengers ($140$~kg, $0.9$~m behind $CM$) and $150$~kg of luggage $0.5$~m behind the rear axle, and their front and rear tires are $12\%$ and $5\%$ softer than nominal. The controllers use nominal unladen parameters. The adaptive law uses $\bar\gamma=0.25$~s$^{-1}$, $\phi_0^2=0.05$, $T_g=1.5$~s, and the set $\Theta$ above. Four configurations are compared:
\begin{itemize}
\item $K_A$, $\alpha=0.5$, with the feedforward of \cite{liu2024robust};
\item $K_A$, $\alpha=0.5$, with the adaptive sideslip-compensated feedforward;
\item $K_B$, $\alpha=1$, with the nominal sideslip-compensated feedforward;
\item $K_B$, $\alpha=1$, with the adaptive feedforward.
\end{itemize}

\begin{figure}[t]
\centering
\includegraphics[width=\columnwidth]{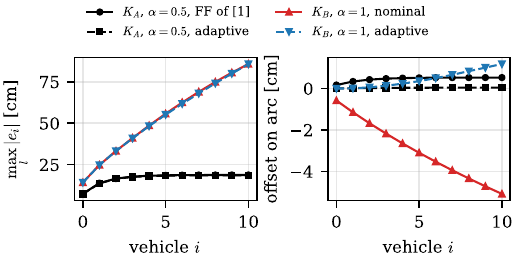}
\caption{Uniformly loaded convoy. Left: largest lateral error of each vehicle. Right: mean offset on the second arc. With $\alpha=1$, errors grow linearly even though $\Hinf{Q_B}=1$ (Theorem~\ref{thm:pred}). With nominal feedforward, the arc offsets accumulate as well. Adaptation removes them.}
\label{fig:convoy}
\end{figure}

\paragraph*{Bounded versus accumulating errors} Fig.~\ref{fig:convoy} (left) shows the main contrast.
\begin{itemize}
\item With $\alpha=0.5$, the largest error rises from $7.4$~cm for the leader to $18.8$~cm and then levels off from about the fifth vehicle, as Theorem~\ref{thm:lp} predicts.
\item With $\alpha=1$, the largest error grows by about $7$~cm per vehicle, reaching $86$~cm at the tenth follower, although $K_B$ satisfies $\Hinf{Q}=1$. The error that each vehicle makes at the curve entries is passed on and added to (Theorem~\ref{thm:pred}). Adaptation does not change this.
\end{itemize}

\paragraph*{Offsets on arcs} With the nominal sideslip-compensated feedforward and $\alpha=1$, the steady offset on the second arc grows linearly from $-0.6$ to $-5.1$~cm. The adaptive law reduces it to between $0$ and $1.2$~cm. The remainder is the slowly decaying tail of the curve-entry transients, which $Q\approx1$ carries forward. With $\alpha=0.5$, the offsets stay bounded, between $0.2$ and $0.5$~cm with the feedforward of \cite{liu2024robust}, and below $0.1$~cm with adaptation.

\begin{figure}[t]
\centering
\includegraphics[width=\columnwidth]{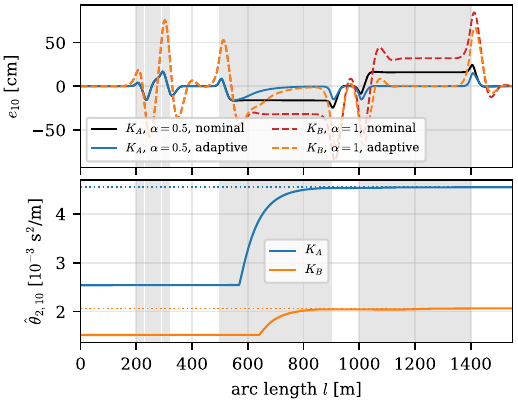}
\caption{Snow and $480$~kg load. Top: lateral error of the last follower. Bottom: its estimate $\hat\theta_2$, with the true values dotted. Shaded: curved segments.}
\label{fig:snow}
\end{figure}

\paragraph*{When adaptation matters} In the dry, moderately loaded convoy above, the feedforward error is small, and adaptation changes the arc offsets by centimetres. The benefit grows with the mismatch. The second scenario is chosen to make it large, and it is also the setting that motivates trail following: snow hides the lane markings and lowers tire stiffness. Every vehicle now carries $480$~kg (two front and two rear passengers, and $200$~kg in the trunk), the cornering stiffnesses are $60\%$ of nominal, and the speed is $20$~m/s with $20$~m gaps. The lateral acceleration on the arcs is $1.6$~m/s$^2$, within the linear range assumed by the model. The true $\theta_2$ is then $1.8$ times the nominal value for $K_A$, and $1.4$ times for $K_B$. Table~\ref{tab:snow} and Fig.~\ref{fig:snow} compare the nominal and the adaptive feedforward. Here, ``nominal'' is the unladen sideslip-compensated feedforward for both gain sets, not the feedforward of \cite{liu2024robust}.
\begin{itemize}
\item With $\alpha=0.5$, the nominal feedforward leaves the last follower $16$~cm off the desired path for the full length of each arc. Adaptation removes this offset, and the RMS error on the arcs falls from $16.0$ to $5.2$~cm. The peak error falls from $24.6$ to $16.9$~cm.
\item With $\alpha=1$, the nominal offsets accumulate to $32$~cm. Adaptation reduces them to $0.6$~cm, and the RMS error on the arcs from $33.9$ to $17.1$~cm. The peak error stays above $70$~cm, because the curve-entry transients still accumulate (Theorem~\ref{thm:pred}).
\end{itemize}
In both cases, $\hat\theta_2$ converges on the first arc to the value given by \eqref{eq:theta} (Fig.~\ref{fig:snow}, bottom) and is reused on the second arc, which has the opposite curvature. In a third run, with randomly drawn loads (0--310~kg, $M_{12}$ from $-438$ to $42$~kg\,m) and tire stiffness between $85\%$ and $105\%$, all estimates converged to within $5\%$ of their true values within about $300$~m of the start of the first arc, at the same rate, as Proposition~\ref{prop:adapt} states.

\begin{table}[t]
\centering
\caption{Snow and $480$~kg load: errors of vehicle $10$ (cm). Offset: mean on the second arc. RMS: over both arcs.}
\label{tab:snow}
\setlength{\tabcolsep}{5pt}
\begin{tabular}{@{}llccc@{}}
\toprule
Configuration & Feedforward & Offset & RMS & Peak\\
\midrule
$K_A$, $\alpha=0.5$ & nominal & 16.0 & 16.0 & 24.6\\
$K_A$, $\alpha=0.5$ & adaptive & 0.0 & 5.2 & 16.9\\
$K_B$, $\alpha=1$ & nominal & 32.2 & 33.9 & 84.7\\
$K_B$, $\alpha=1$ & adaptive & 0.6 & 17.1 & 75.6\\
\bottomrule
\end{tabular}
\end{table}

\section{Conclusions}

For convoys that steer along GNSS trails, the analysis rests on three modeling choices:
\begin{itemize}
\item the heading reference is the trail's course angle;
\item errors are compared at the same arc length;
\item the curvature feedforward on the fitted trail is part of the propagation.
\end{itemize}
With these choices, the propagation is scalar. Leader-plus-predecessor information yields errors that are bounded uniformly in convoy length if (and, for identical vehicles, only if) $\alpha\Hinf{Q}<1$, and the design of \cite{liu2024robust} meets this condition with margin. Predecessor-only following can be made non-amplifying with exact feedforward, but $Q(0)=1$ makes the errors injected by each vehicle accumulate. Loads and tire stiffness enter the feedforward through one coefficient. Estimating it on arcs with a gated, projected law removes the resulting offsets, and the averaged estimator dynamics are decoupled across the convoy. The benefit is a few centimetres for a lightly loaded car on dry pavement, and becomes substantial when the tires are soft and the vehicles heavily loaded, as on snow. Varying speed, measurement noise in the fitted curvature, and an experimental check on instrumented vehicles are left for future work.

\bibliographystyle{IEEEtran}
\bibliography{refs}

\end{document}